\documentclass[showpacs,preprintnumbers]{revtex4}
\usepackage{graphicx}

\begin{document}

\date{\today }
\title{Trions in coupled quantum wells and Wigner crystallization}
\author{Oleg L. Berman$^{1,2}$, Roman Ya. Kezerashvili$^{1,2}$, and Shalva
M. Tsiklauri$^{1}$}
\affiliation{\mbox{$^{1}$Physics Department, New York
City College of Technology, The City University of New York,} \\
Brooklyn, NY 11201, USA \\
\mbox{$^{2}$The Graduate School and University Center, The
City University of New York,} \\
New York, NY 10016, USA }

\begin{abstract}
We consider a restricted three body problem, where two interacted particles
are located in two dimensional (2D) plane and interact with the third one
located in the parallel spatially separated plane. The system of such type
can be formed in the semiconductor coupled quantum wells, where the
electrons (or holes) and direct excitons spatially separated in different
parallel neighboring quantum wells that are sufficiently close to interact
and form negative $X^{-}$ or positive $X^{+}$ indirect trions. It is shown
that at large interwell separations, when the interwell separation much
greater than the exciton Bohr radius, this problem can be solved
analytically using the cluster approach. Analytical results for the energy
spectrum and the wave functions of the spatially indirect trion are
obtained, their dependence on the interwell separations is analyzed and a
conditional probability distribution is calculated. The formation of 2D
Wigner crystal of trions at the low densities is predicted. It is shown that
the critical density of the formation of the trion Wigner crystal is
sufficiently greater than the critical density of the electron Wigner
crystal.
\end{abstract}

\pacs{71.35.Pq, 71.35.-y, 73.21.Fg}
\maketitle





\section{Introduction}

\label{intro}

The linear and nonlinear properties of semiconductor heterostructures are
often governed by excitons that are defined as the bounded state of the
electron-hole pair. This is especially important for semiconductors with
wide-bandgap, where the excitonic binding energy is comparable to room
temperature. Positively ($X^{+}$) and negatively ($X^{-}$) charged excitons,
called trions in semiconductor structures, have been the subject of many
experimental and theoretical studies in the last years. Positively charges
excitons are formed by one electron and two holes, and negatively charged
excitons are formed by two electrons and one hole. The theoretical proof of
the stability of trions in bulk semiconductors was presented by Lampert~\cite%
{Lampert}. The experimental observation of trions in semiconductor quantum
wells (QWs) was achieved in CdTe/CdZnTe~\cite{Kheng} and in GaAs/AlGaAs~\cite%
{Finkelstein, Shields_2, Shields}.

The many-body formalism including Feynman diagram technique was developed to
study the collective properties of three-dimensional (3D) trions in the bulk
semiconductors~\cite{Combescot}. The 3D $X^{+}$ and $X^{-}$ confined in a
semiconductor cylindrical quantum dot were investigated using a variational
procedure within the effective mass approximation~\cite{Safwan}. The 3D
trions in bulk semiconductors~\cite{Munchy} and two-dimensional trions~\cite%
{Stebe} were studied theoretically. According to these studies, the 2D
trions have binding energies that are larger than the trions in the
corresponding bulk materials. The 2D excitons and 2D $X^{-}$ in QWs in the
presence of 2D electron gas were observed experimentally by optical
measurements of an excitonic recombination line on the photoluminescence
spectra from the QW~\cite{Finkelstein_2}. The experiments devoted to the
observation of 2D $X^{-}$ in QWs in magnetic field by studying the
magneto-optical absorption spectra were performed in Ref.~[\onlinecite{Kheng}%
]. Besides, the 2D $X^{-}$ in QWs in magnetic field were observed as a
results of the analysis of the excitonic recombination line on the
electroluminescence spectra from the QW~\cite{Buhmann}. The experiments
devoted to the observation of magneto-optical spectra of 2D $X^{-}$ in QWs
were performed~\cite{Shields_1,Shields_2}. In Ref.~[\onlinecite{Shields_2}]
the 2D $X^{-}$ are studied in coupled quantum wells, where electrons and
holes are spatially separated in different QWs.

The problem of finding the wave functions and eigenenergies of $X^{-}$ and $%
X^{+}$ formed by three particles is the tree-body problem. There are
different approaches were applied to solve three-body problem: the numerical
diagonalization of Hamiltonian for the eigenvalue problem, the variational
approach, the Hartree-Fock method, the approach of Faddeev equations, the
method of the hyperspherical functions (HF). The three-electron quantum dot
in a harmonic confinement potential was studied using the HF method~\cite%
{Johnson,Ruan} and Faddeev equations in the configuration space~\cite{Braun}%
. The hyperspherical functions method was applied to solve the Schr\"{o}%
dinger equation for 2D $X^{+}$ and $X^{-}$~\cite{Ruan_2}. The three-body
problem for three electrons in a quantum dot in a magnetic field was solved
in the framework of the hyperspherical functions method~\cite{Kezerashvili}.
The variational method was applied to solve the eigenvalues and
eigenfucntions problem for 3D $X^{+}$ and $X^{-}$ in cylindrical quantum dot~%
\cite{Safwan}. A variational calculation of the ground-state energy of 2D
neutral excitons, $X^{+}$, and $X^{-}$ in a single QW was performed~in Ref. [%
\onlinecite{Riva}]. A variational calculation of the lower singlet and
triplet states of positively charged 2D trions, confined to a single quantum
well and in the presence of a perpendicular magnetic field, was performed~%
\cite{Riva2}. The ground-state energy of 2D $X^{+}$ in coupled QWs (CQWs)
formed by spatially separated two 2D heavy holes in one QW and one 2D
electron in neighboring QW was calculated using the variational method
within the infinite-hole-mass approximation~\cite{Sergeev}. The 2D $X^{-}$
in the presence of perpendicular magnetic field were studied numerically by
the diagonalization of the Hamiltonian in Refs.~[%
\onlinecite{Chapman,Whittaker}]. The ground state energy of the 2D $X^{-}$
in the presence of 2D electron gas was calculated using the diagonalization
of the eigenvalue problem in the random phase approximation~\cite{Dacal}. In
Ref.~[\onlinecite{Dacal}] the screening of the Coulomb interaction and the
Pauli exclusion principle were considered. A first-principle path integral
Monte Carlo study of the binding energy of excitons, 2D positively and
negatively charged trions and biexcitons bound to single-island interface
defects in 2D GaAs/AlGaAs quantum wells was presented in Ref.~[%
\onlinecite{Filinov}]. The preliminary results for trions in the system of
semiconductor coupled quantum wells were reported at the 21 European
Conference on few-body problem in physics [\onlinecite{bk}].

In all theoretical studies mentioned above the Schr\"{o}dinger equation for
the trion was solved numerically using the different methods. In this Paper
we propose the type of the trion which wave functions and eigenenergies can
be obtained analytically for some conditions. We consider two parallel CQWs.
In the case 1 there are electrons in one QW and the electrons and holes in
the other QW. In the case 2 there are holes in one QW and the electrons and
holes in the other QW. In the case 1 we consider $X^{-}$ formed by one
electron in one QW and an electron and a hole in the other QW. In the case 2
we consider $X^{+}$ formed by one hole in one QW and an electron and a hole
in the other QW. The eigenfunction and eigenvalue problem for these trions
is the restricted three-body problem, since the motion of the electron (or
hole) is restricted in the plane of one QW, and the motion of the
electron-hole pair is restricted in the plane of the other QW. We show that
for relatively large distance between these CQWs the eigenvalue and
eigenfucntion problem for this trions can be solved analytically.

The dilute system of trions at the densities, when the average distance
between the trions is much larger than the radius of each trion the system
of trions can be treated as the dilute system of the fermions with the pair
Coulomb repulsion, which can form non-electronic Wigner crystal. The 3D
electron Wigner crystal was described in Ref.~[\onlinecite{Pines}] and the
2D electron Wigner crystal was described in Ref.~[\onlinecite{Maradudin}].
The 2D electron Wigner crystal formed by electrons on films of helium was
studied in Refs.~[\onlinecite{Peeters_Platzman,Peeters}]. The spectrum of
collective excitations for the 2D electron Wigner crystal formed by
electrons on a helium film was calculated~\cite{Peeters}. The ground state
of a 2D electron gas including formation of the 2D Wigner crystal at low
densities was calculated using Monte-Carlo method~\cite{Tanatar}. Wigner
crystallization of electrons in 2D quantum dots was analyzed using path
integral Monte Carlo approach~\cite{Filinov2}. We predict the formation of
2D Wigner crystal of trions at the low densities. In our consideration we
neglect the influence of the disorder in the quantum wells. We will show
that the critical density of the formation of the trion Wigner crystal is
sufficiently larger than the critical density of the electron Wigner
crystal, since the mass of trion is larger than the mass of the electron.

The Paper is organized in the following way. In Sec.~\ref{Hamiltonian} we
introduce the model Hamiltonian for $X^{-}$ and $X^{+}$ in the CQWs reducing
the restricted three-body problem to the 2D three-body problem for the
exciton and the projection of the electron or hole on the plane of the
excitonic QW.  In Sec.~\ref{exciton} we review the solution of the Schr\"{o}%
dinger for 2D exciton in a single QW. In Sec.~\ref{el-ex} we find the wave
function of the relative motion of the exciton and electron (hole) and
obtain the corresponding energy spectrum. The energy spectrum for the trion,
it dependence on the interwell separation and calculation of the conditional
probability distribution based on the trion wave functions are presented in
Sec.~\ref{disc}. The formation of the trion Wigner crystal in the dilute
trion gas is discussed in Sec.~\ref{Wig}. Finally, our  conclusions follow
in Sec.~\ref{conc}.

\section{The total Hamiltonian and wave function for an isolated trion}

\label{Hamiltonian}

We consider two parallel coupled quantum wells separated by a spatial
barrier of thickness $D$. There are excitons in one QW and electrons or
holes in the other QW. One method for obtaining a system of excitons is to
create excitons in one region of the QW using laser pumping. 2D electrons or
holes can be obtained in the QW by doping semiconductors that tends to
create a constant Fermi energy, producing an excess of carriers, either
electrons (donors) or holes (acceptors). These excess carriers then give the
material either electrons in the conduction bands or holes in the valence
band.

Let $r_{i}$ be the coordinates of particles in laboratory frame with
effective masses of the electron $m_{e}$ and the hole $m_{h}$. Let's
introduce Jacobi coordinates which are an appropriate set for the three-body
problem:

\begin{eqnarray}
\mathbf{\eta } &=&(\mathbf{r}_{2}-\mathbf{r}_{1})  \nonumber \\
\mathbf{r}_{e(h)} &=&\left( \mathbf{r}_{3}-\frac{m_{e}\mathbf{r}_{1}+m_{h}%
\mathbf{r}_{2}}{m_{e}+m_{h}}\right)  \label{rk0} \\
\mathbf{R} &\mathbf{=}&\frac{1}{M}\left( m_{e}\mathbf{r}_{1}+m_{h}\mathbf{r}%
_{2}+m_{e(h)}\mathbf{r}_{3}\right)  \nonumber
\end{eqnarray}%
where $m_{e(h)}$ is the electron (hole) mass, the coordinate vector $\mathbf{%
\eta }$ in 2D space describes of the relative motion of the electron and
hole in the exciton with the reduce mass $\mu =m_{e}m_{h}/(m_{e}+m_{h})$ in
the excitonic QW$,$ the coordinate vector $\mathbf{r}_{e(h)}$ from the
center of mass of the exciton to the electron (hole) in the other QW
describes the motion of the particles with reduce mass $m=$ $%
(m_{e}+m_{h})m_{e(h)}/(m_{e}+m_{h}+m_{e(h)}),$ $R$ is the vector of the
center of mass of trion and $M=m_{e}+m_{h}+m_{e(h)}$ is the mass of a
negatively or positively charged trion.

The total Hamiltonian for an exciton in one QW interacted with an electron
(hole) in the other QW is given by
\begin{equation}
\hat{H}=-\frac{\hbar ^{2}}{2m_{e(h)}}\Delta _{e(h)}+\hat{H}%
_{ex}+V_{e(h)-ex}\ ,  \label{rk1}
\end{equation}%
%
%
%
%
%
%
%
where $V_{e(h)-ex}$ is the potential energy of the interaction between the
electron (hole) in one QW and the exciton in an excitonic QW, and $\hat{H}%
_{ex}$ is the exciton Hamiltonian.

\begin{figure}[tbp]
\includegraphics[width = 3.5in]{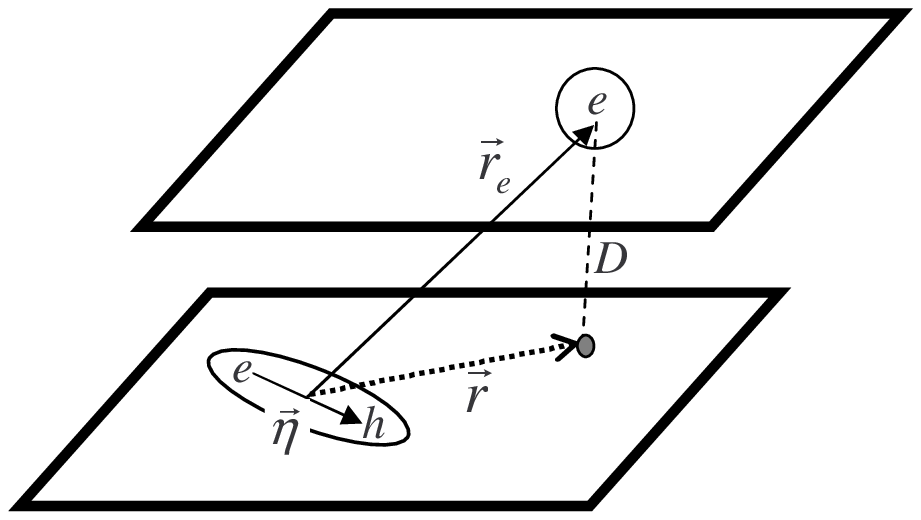}
\caption{The restricted three-body problem in the CQWs: an electron in one
QW and an electron-hole pair in the other QW.}
\label{restr}
\end{figure}

Since the motion of the exciton is restricted in the excitonic QW and the
motion of the electron (hole) is restricted in the electronic QW, we replace
the coordinate vector of the electron (hole) $\mathbf{r}_{e(h)}$ by its
projection $\mathbf{r}$ on plane of the excitonic QW using the relation $%
r_{e(h)}=\left( r^{2}+D^{2}\right) ^{1/2}.$ Thus, we reduced the restricted
3D three-body problem to the 2D three-body problem on the plane of the
excitonic QW as shown in Fig.~\ref{restr}.

The Hamiltonian of the electron and the hole forming the exciton in the
excitonic QW has the following form

\begin{equation}
\hat{H}_{ex}=-\frac{\hbar ^{2}}{2m_{e}}\Delta _{\mathbf{r}_{1}}-\frac{\hbar
^{2}}{2m_{h}}\Delta _{\mathbf{r}_{2}}-\frac{e^{2}}{\epsilon \eta }\ ,
\label{rk2}
\end{equation}%
where $\Delta _{\mathbf{r}_{1}}$ and $\Delta _{\mathbf{r}_{2}}$ are the
Laplacian operators with respect to the components of the vectors $\mathbf{r}%
_{1}$ and $\mathbf{r}_{2}$ respectively, $\epsilon $ is the dielectric
constant, $e$ is the electron charge. The problem of the in-plane motion of
interacting electron and hole with effective masses $m_{e}$ and $m_{h}$,
respectively, forming the exciton in the excitonic QW can be reduced to that
of one particle with reduced mass $\mu $ and energy $E_{ex}$, moving in a
Coulomb potential and motion of the center of mass of the exciton.

Let us assume, that the distance between QWs $D$ is greater than the Bohr
radius $a_{B}$ of the exciton formed by the bounded state of the electron
and the hole in the excitonic QW: $D\gg a_{B}$, where $a_{B}=\hbar
^{2}\epsilon /(2m_{ex}e^{2})$. Under this assumption the potential energy $%
V_{e(h)-ex}$ of the pair interactions between the exciton and electron
placed in two parallel quantum wells which is the sum of the Coulomb
electron-electron repulsion and electron-hole attraction for the $X^{-}$
trion or the hole-hole repulsion and electron-hole attraction for the $X^{+}$
trion can be reduced to the polarization energy. Follow Ref. \cite{Nikitkov}
the polarization energy for the interaction of the isolated 2D exciton and
an isolated electron can be written in an approximation in the form

\bigskip
\begin{equation}
V_{e(h)-ex}(r)=-\frac{\alpha }{r_{e(h)}^{4}}=-\frac{\alpha }{%
(r^{2}+D^{2})^{2}}\ ,  \label{rk3}
\end{equation}%
%
%
%
%
%
%
%
%
%
%
%
where $\alpha =\frac{21}{32}\frac{e^{2}a_{B}^{3}}{\epsilon }\ $\ and $r$ is
the distance between the center of mass of the exciton and the projection of
the electron (hole) on the excitonic plane as shown in Fig.~\ref{restr}. In
other words, $\mathbf{r}$ is the projection of the coordinate $\mathbf{%
r_{e(h)}}$ defined by Eq.~(\ref{rk0}) onto the excitonic plane.


\bigskip Based on Eqs. (\ref{rk1}) - (\ref{rk3}) after the separation of the
motion of the center of mass of the trion the eigenstates of the trion in
coordinates (\ref{rk0}) are described by the following Schr\"{o}dinger
equation%
\begin{equation}
\left( -\frac{\hbar ^{2}}{2\mu }\Delta _{\mathbf{\eta }}-\frac{e^{2}}{%
\epsilon \eta }-\frac{\hbar ^{2}}{2m}\Delta _{\mathbf{r}}-\frac{\alpha }{%
(r^{2}+D^{2})^{2}}\right) \Psi (\mathbf{\eta ,r)=}\ E\ \Psi (\mathbf{\eta ,r)%
} \ ,  \label{rk4}
\end{equation}%
where $\Delta _{\mathbf{\eta }}$ and $\Delta _{\mathbf{r}}$ are the
Laplacian operators with respect to the components of the vectors $\mathbf{%
\eta ,}$ and $\mathbf{r}$, respectively. Eq. (\ref{rk4}) allows the
separation of variables and reduces to the Schr\"{o}dinger equation that
describes the eigenstates of a 2D exciton, and the Schr\"{o}dinger equation
that describes the relative motion of the exciton and electron or exciton
and hole in the 2D excitonic plane.

Seeking the solution of (\ref{rk4}) in the form $\Psi (\mathbf{\eta ,r)=}%
\Psi _{ex}(\mathbf{\eta })\Psi _{e(h)}(\mathbf{r})$ and using the separation
of variables we obtain%
\begin{equation}
\left( -\frac{\hbar ^{2}}{2\mu }\Delta _{\mathbf{\eta }}-\frac{e^{2}}{%
\epsilon \eta }\right) \Psi _{ex}(\mathbf{\eta )=}\ E_{ex}\ \Psi _{ex}(%
\mathbf{\eta )},  \label{rk5}
\end{equation}

\begin{equation}
\left( -\frac{\hbar ^{2}}{2m}\Delta _{\mathbf{r}}-\frac{\alpha }{%
(r^{2}+D^{2})^{2}}\right) \Psi _{e(h)}(\mathbf{r)=}\ E_{e(h)}\ \Psi _{e(h)}(%
\mathbf{r)}\ .  \label{rk6}
\end{equation}

\bigskip Therefore, the energy spectrum of the trion is
\begin{equation}
E=E_{ex}+E_{e(h)}.  \label{rk7}
\end{equation}

Let us consider (\ref{rk5}) and (\ref{rk6}) separately.

\section{The wave function and energy spectrum of the 2D exciton}

\label{exciton}

The eigenstates of the electron and the hole forming a 2D exciton in
the excitonic QW are described by the Schr\"{o}dinger (\ref{rk5}),
that represents a two-body problem which is like the two-dimensional
hydrogenic problem.  Below we follow the approach of Ref.
\cite{Zaslow} for finding the wavefunctions and eigenenergies of the
2D hydrogen atom in polar coordinates for the arbitrary values of
two masses. The Schr\"{o}dinger equation (\ref{rk5}) that describes
the relative 2D motion of an electron
and hole, in polar coordinates has the form%
\begin{equation}
\left[ -\frac{\hbar ^{2}}{2\mu }\left( \frac{\partial ^{2}}{\partial \eta
^{2}}+\frac{1}{\eta }\frac{\partial }{\partial \eta }+\frac{1}{\eta ^{2}}%
\frac{\partial ^{2}}{\partial \varphi ^{2}}\right) -\frac{e^{2}}{\epsilon
\eta }\right] \Psi _{ex}(\eta ,\varphi \mathbf{)=}\ E_{ex}\ \Psi _{ex}(\eta
,\varphi \mathbf{)}.  \label{rk8}
\end{equation}%
After the standard separation of the variables $\eta $ and $\varphi
$ we obtain the radial Schr\"{o}dinger equation and the solution for
the 2D exciton wave function in terms of associated Laguerre
polynomials is presented in \cite{Zaslow}, and the exciton spectrum
is given by
\begin{equation}
E_{n} \equiv E_{ex} =-\frac{\mu e^{4}}{2\epsilon
^{2}(n-1/2)^{2}\hbar ^{2}}\ .  \label{rk10}
\end{equation}%
where  $n=1,2,3,\ldots $ are the quantum numbers.
We can see that in the ground state at $n=1$ the binding energy of the
exciton is $E_{1}=2\mu e^{4}/(\epsilon ^{2}\hbar ^{2})$, and the
characteristic radius of the exciton corresponding to the wave function $%
\Psi _{ex}(\mathbf{\eta )}$ at $n=1,$ $l=0$ is given by $a_{B}=\hbar
^{2}\epsilon /(2\mu e^{2})$. Interestingly enough that by replacing $(n-1/2)$
by $n$ retrieves relevant formulas of the 3D case, particularly quantization
conditions (\ref{rk10}).


\bigskip

\section{The wave function and energy of the relative motion of the exciton
and electron and exciton and hole}

\label{el-ex}

The eigenstates of the electron (hole) in the field of the exciton are
described by the Schr\"{o}dinger (\ref{rk6}), which represents a motion of
the projection of the electron (hole) on the plane of the excitonic QW.
Using the polar coordinates $\mathbf{r}$ and $\phi$ and substituting $\Psi
_{e(h)}(r,\phi ) = r^{-1/2}u_{e(h)}(r,\phi)$ into Eq.~(\ref{rk6}), we obtain
the 2D Schr\"{o}dinger equation of the relative motion of an exciton and an
electron (hole)
\begin{eqnarray}
\left[- \frac{\hbar ^{2}}{2m}\left( \frac{\partial ^{2}}{\partial r^{2}}+
\frac{1}{4r^{2}} +\frac{1}{r^{2}}\frac{\partial ^{2}}{\partial \phi ^{2}}
\right) + V_{e(h)-ex}(r) \right] u_{e(h)}(r,\phi) = E_{ex} u_{e(h)}(r,\phi)
\ ,  \label{rk121}
\end{eqnarray}
where $V_{e(h)-ex}(r)$ is given by Eq.~(\ref{rk3}). The 2D motion of the
projection of the electron (hole) in the plane of the excitonic QW, can be
treated as that of a single particle with reduced mass $m$ and energy $%
E_{e(h)}$, moving in a potential $\frac{\alpha }{(r^{2}+D^{2})^{2}}.$
Assuming $r\ll D$, the Taylor expansion of $V_{e(h)-ex}(r)$ from Eq.~(\ref%
{rk3}) up to the linear term results in
\begin{equation}
V_{e(h)-ex}(r)=-\frac{\alpha }{D^{4}}+\frac{1}{2}\frac{\alpha }{D^{6}}\
r^{2}.  \label{rk11}
\end{equation}

Using potential (\ref{rk11}) in (\ref{rk121}) we obtain the Schr\"{o}%
dinger's equation that describes the two-dimensional motion of a
single electron (hole), that  corresponds to the 2D harmonic
oscillator with the harmonic parabolic potential $\gamma
\frac{r^{2}}{2}$, where $\gamma =\alpha /D^{6}$, when the energy is
counted from the energy level $E_{e(h)}^{(0}=-\alpha /D^{4}$.

Using the standard separation of the variables $r$ and $\phi $ in  and follow Refs.~[\onlinecite{Maksym,Iyengar}] we obtain the radial Schr%
\"{o}dinger equation and the solution for the eigenfunctions for the
projection of the electron (hole) on the plane of the excitonic QW
can be written in terms of associated Laguerre polynomials. The characteristic radius $a=\left( \hbar /\left( 2\sqrt{m\gamma }\right) \right) ^{1/2}$ of
the wave function  is directly proportional to $D^{3/2}$ ($%
a\sim a_{B}^{-1/2}D^{3/2}$). The corresponding energy spectrum is
given by
\begin{equation}
E_{N\text{ }L} \equiv E_{e(h)} = - \frac{\alpha}{D^{4}} +
(2N+1+|L|)\hbar \left( \frac{\alpha}{D^{6}m}\right) ^{1/2}\ .
\label{rk15}
\end{equation}%
where $N=\mathrm{min}(\widetilde{n},\widetilde{n}^{\prime })$, $L=\widetilde{%
n}-\widetilde{n}^{\prime }$, $\widetilde{n},$ $\widetilde{n}^{\prime
}=0,1,2,3,\ldots $ are the quantum numbers.

As it seen from Eq.~(\ref{rk15}), the first term is proportional to $D^{-4}$%
, and the second term is proportional to $D^{-3}$. Therefore, for the some
values of the separation distance $D$ between the QW the energy becomes
positive. That means that at the interwell separations greater than some
critical $D_{cr}$, $D > D_{cr}$ the energy $E_{N\text{ }L}$ is positive, $%
E_{N\text{ }L} > 0$, and at $D < D_{cr}$, the energy $E_{N\text{ }L}$ is
negative, $E_{N\text{ }L} < 0$. For example, at the lowest quantum state $N
= L = 0$ as it follows from Eq.~(\ref{rk15}) the ground state energy for the
electron (hole) in the parabolic well of the exciton in the excitonic QW is
given by
\begin{equation}
E_{00}=-\frac{\alpha }{D^{4}}+\hbar \left( \frac{\alpha }{D^{6}m}\right)
^{1/2}\ \ .  \label{rk16}
\end{equation}

Using Eq.~(\ref{rk16}), and the value for $\alpha $ we obtain the critical
interwell separation $D_{cr}$:
\begin{equation}
D_{cr}=\left( \frac{21}{32}\right) ^{1/2}\frac{ea_{B}^{3/2}m^{1/2}}{\hbar
\epsilon ^{1/2}}\ .  \label{rk17}
\end{equation}%
%
%
%
%
%
%
%
%
%
If we consider the trions in GaAs/AlGaAs CQWs, then we have $m_{e}=0.07m_{0}$%
, $m_{h}=0.15m_{0}$, $\epsilon =13$, where $m_{0}$ is the free electron
mass. Using Eq.~(\ref{rk17}) for the GaAs/AlGaAs CQWs we obtain for $X^{-}$
the critical interwell separation $D_{cr}=4.4\ \mathrm{nm}$, and for $X^{+}$
we find $D_{cr}=5.6\ \mathrm{nm}$. However, let us mention that our approach
is valid only for $D\gg D_{cr}$, because the condition of the Taylor
expansion of the electron (hole)-exciton interaction potential energy is $%
D\gg a_{B}$.

\section{Discussion of the results for the single trion}

\label{disc}

In our cluster approach, when the trion is the system of interacting exciton
and electron or hole, the wavefunction of the trion $\Psi (\mathbf{\eta },%
\mathbf{r})$ is a product of the wavefunctions of the exciton and
relative motion of exciton and electron (hole) and, therefore, can
be expressed in terms of associated Laguerre polynomials. As it
follows from (\ref{rk7}) the energy spectrum of the trion is defined
by Eqs.~(\ref{rk10}) and (\ref{rk15}) and is given by

\begin{equation}
E_{n\text{ }N\text{ }L}=-\frac{21}{32}\frac{e^{2}a_{B}^{3}}{\epsilon D^{4}}-%
\frac{\mu e^{4}}{2\epsilon ^{2}(n-1/2)^{2}\hbar ^{2}}+(2N+1+|L|)\hbar \left(
\frac{21}{64}\frac{e^{2}a_{B}^{3}}{\epsilon mD^{6}}\right) ^{1/2}.
\label{rk18}
\end{equation}

\begin{figure}[tbp]
\includegraphics[width = 4.0in]{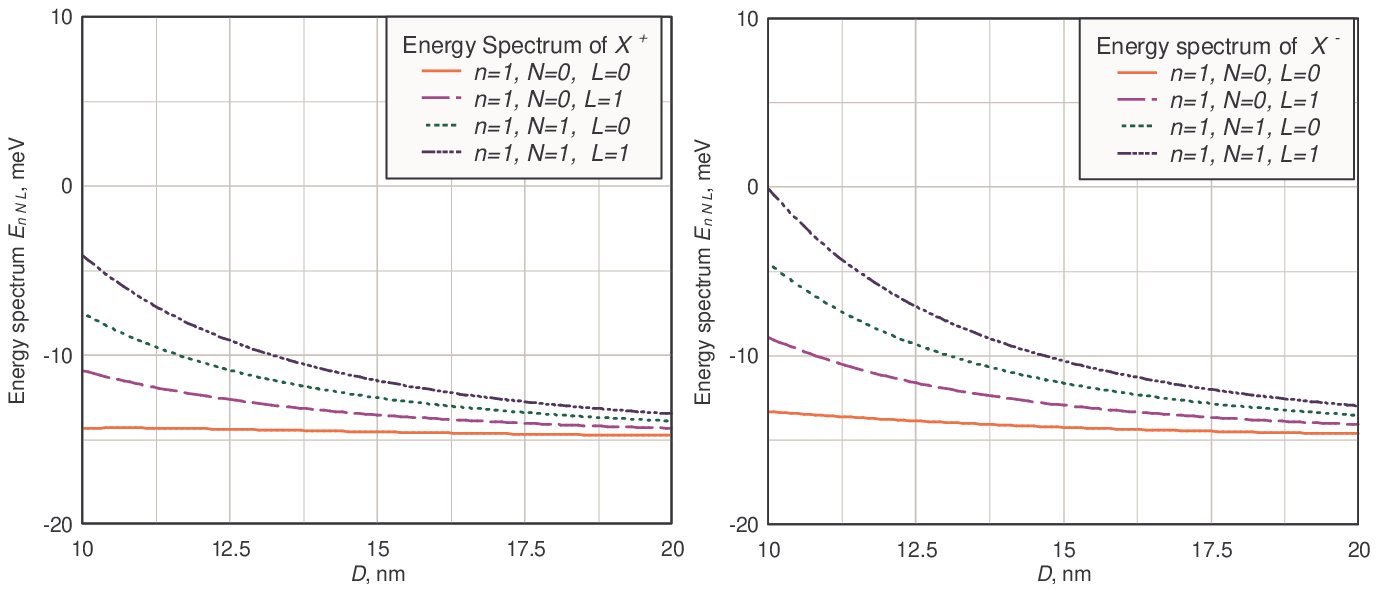}
\caption{The energy spectrum for $X^{+}$ and $X^{-}$ as a function of the
interwell separation $D$ for different quantum states.}
\label{x}
\end{figure}

The energy spectrum of the trion corresponds to the discrete levels
corresponding to the bounded states of the electron (or hole) and exciton.
Therefore, the trion wavefunction represents the bounded state with the
characteristic radius at large $D$ corresponding to the radius of the trion $%
a_{t}=a$. Since $a$ is directly proportional to $D^{3/2}$, which is the
highest degree of $D$ compare to the other characteristic radiuses, thus the
radius of the trion is proportional to $D^{3/2}$. The calculated energy
spectrum as a function of the interwell separation $D$ for different quantum
numbers for $X^{+}$ and $X^{-}$ are shown in Fig.~\ref{x}.

\begin{table}[ht]
\caption{The energy spectrum of $X^{+}$ and $X^{-}$ trions for interwell
separation $D=12$ nm for different quantum states.}\label{table1}
\par
\begin{center}
\bigskip
\begin{tabular}{|cccc||cccc|}
\hline
$n$ & $N$ & $L$ & $X^{+},$ Energy, meV & $n$ & $N$ & $L$ & $X^{-},$ Energy,
meV \\ \hline
1 & 0 & 0 & \ \ \ \ -14.7 & 1 & 0 & 0 & \ \ \ \ -14.1 \\
1 & 0 & 1 & \ \ \ \ -12.7 & 1 & 0 & 1 & \ \ \ \ -11.6 \\
1 & 0 & 2 & \ \ \ \ -10.8 & 1 & 0 & 2 & \ \ \ \ \ -9.0 \\
1 & 0 & 3 & \ \ \ \ \ -8.8 & 1 & 0 & 3 & \ \ \ \ \ \ -6.5 \\
1 & 0 & 4 & \ \ \ \ \ -6.8 & 1 & 0 & 4 & \ \ \ \ \ \ -3.9 \\
1 & 0 & 5 & \ \ \ \ \ -4.8 & 1 & 0 & 5 & \ \ \ \ \ \ -1.3 \\
1 & 0 & 6 & \ \ \ \ \ -2.9 &  &  &  &  \\
1 & 0 & 7 & \ \ \ \ \ -0.9 &  &  &  &  \\
&  &  &  &  &  &  &  \\
1 & 1 & 0 & \ \ \ \ -10.8 & 1 & 1 & 0 & \ \ \ \ \ \ -9.8 \\
1 & 1 & 1 & \ \ \ \ \ -8.8 & 1 & 1 & 1 & \ \ \ \ \ \ -6.5 \\
1 & 1 & 2 & \ \ \ \ \ -6.8 & 1 & 1 & 2 & \ \ \ \ \ \ -3.9 \\
1 & 1 & 3 & \ \ \ \ \ -4.8 & 1 & 1 & 3 & \ \ \ \ \ \ -1.3 \\
1 & 1 & 4 & \ \ \ \ \ -2.8 &  &  &  &  \\
1 & 1 & 5 & \ \ \ \ \ -0.9 &  &  &  &  \\
&  &  &  &  &  &  &  \\
1 & 2 & 0 & \ \ \ \ -6.8 & 1 & 2 & 0 & \ \ \ \ \ \ -3.9 \\
1 & 2 & 1 & \ \ \ \ -4.8 & 1 & 2 & 1 & \ \ \ \ \ \ -1.3 \\
1 & 2 & 2 & \ \ \ \ -2.9 &  &  &  &  \\
1 & 2 & 3 & \ \ \ \ -0.9 &  &  &  &  \\
&  &  &  &  &  &  &  \\
1 & 3 & 0 & \ \ \ \ \ -2.9 &  &  &  &  \\
1 & 3 & 1 & \ \ \ \ \ -0.9 &  &  &  &  \\
&  &  &  &  &  &  &  \\
2 & 0 & 0 & \ \ \ \ \ -1.0 & 2 & 0 & 0 & \ \ \ \ \ \ -0.5%
\end{tabular}%
\ \ \ \ \ \ \ \ \ \ \ \ \ \ \ \ \ \ \ \ \ \ \ \ \ \ \ \ \ \ \ \
\end{center}
\end{table}

Eq.~(\ref{rk18}) shows the dependence of the energy spectrum of a trion on
the interwell separation $D$. Typical thickness of the barrier between two
parallel QWs $D$ varies from one experiment to the other experiment. The
thickness of barrier separation for production of indirect excitons was $4 \
\mathrm{nm}$~\cite{Butov1,Butov2}. However, for interwell distances used in
the drag experiments in Ref.~[\onlinecite{Gramila}] are $D = 17.5 \ \mathrm{%
nm}$ and $D = 22.5 \ \mathrm{nm}$, and the interwell separations used in the
drag experiment in Ref.~[\onlinecite{Lilly2}] are $20 \ \mathrm{nm}$ and $30
\ \mathrm{nm}$. Therefore, we can conclude that it is possible to
experimentally construct the QWs with thickness of the barrier from $4 \
\mathrm{nm}$ to $30 \ \mathrm{nm}$. In our calculations we vary the
interwell separation within these limits (from $10 \ \mathrm{nm}$ to $20 \
\mathrm{nm}$).  Analysis of the energy spectrum of the trion, for example
for $D=12$ nm, shows that $X^{+}$ exists when the exciton is in the ground
state with $n=1$ and the hole is in states: $N=0,\ L=0,1,\ldots ,7$ or the
exciton is in state $n=2$ and the hole is in the ground state. $X^{-}$
exists when the exciton is in the ground state and the electron is in
states: $N=0,\ L=0,1,\ldots ,5$ or the exciton is in the state $n=2$ and the
hole is in the ground state. Let us once again mention that there is the
critical interwell separations when for $D\geq D_{cr}$ the the ground state
energy of the trion is negative, and if we consider the trions in
GaAs/AlGaAs coupled QWs, then we obtain for $X^{-}$ the critical interwell
separation $D_{cr}=4.4\ \mathrm{nm}$, and for $X^{+}$ we find $D_{cr}=5.6\
\mathrm{nm}$. However, our approach is valid only for $D\gg D_{cr}$, because
the condition of the Taylor expansion of the electron (hole)-exciton
interaction potential energy is $D\gg a_{B}$.

\bigskip It is interesting to mention that from Eq. (\ref{rk18}) follows
that, for example, for $L=0$ the maximum value of $N=3$ for distance $D=12$
nm and $N=7$ for $D=15$ nm when the exciton is in ground state and the
energy of the trion is still negative. The energies spectrum of $X^{+}$ and $%
X^{-}$ trions for the interwell separation $D=12$ nm for different quantum
numbers are presented in the Table~I.

Analysis of Table I shows that the $X^{-}$ trion does not exist in the
states $(n,N,L) = (1,0,6), \ (1,0,7), \ (1, 1, 4), \ (1,1,5)), \ (1,2,2), \
(1,2,3), \ (1,3,0), \ (1,3,1)$, while the the $X^{+}$ trion exists in all
these states. It is clear from Eq.~(\ref{rk18}) that the energy spectrum
depends on the trion reduce mass $m$. The reduced  mass for the $X^{+}$
trion is $0.89 m_{0}$ and it is greater than one for the $X^{-}$ trion,
which is $0.53 m_{0}$,  because the effective hole mass $m_{h}$ is greater
than the effective electron mass $m_{e}$, and, therefore, the second
positive term for the trion energy in Eq.~(\ref{rk18}) is greater for $X^{-}$
than for $X^{+}$, since the second term is proportional to $m^{-1/2}$.
Therefore, for the states listed above, the binding energy for $X^{-}$ is
positive, and the bound state for $X^{-}$ does not exist, while for the same
quantum states the binding energy for $X^{+}$ is negative, and the bound
state for $X^{+}$ exists.

\begin{figure}[tbp]
\includegraphics[width = 4.0in]{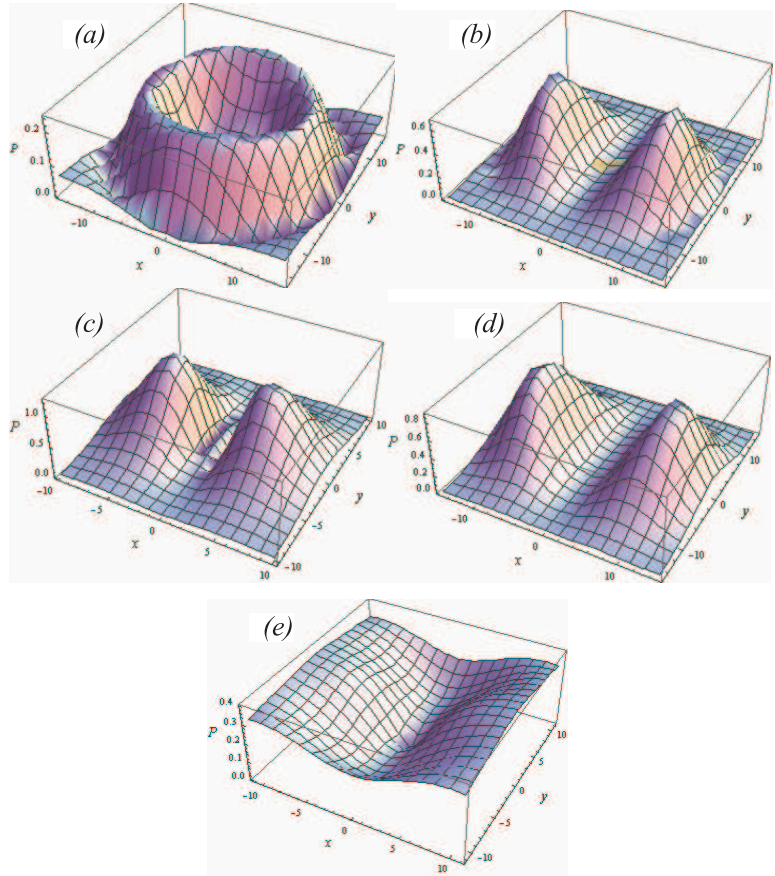}
\caption{ The CPDs for $X^{+}$ trion for the state $n=1$, $l=0$, $N=2$, $L=0$
and different inrewell separations: (a) $D=10 \ \mathrm{nm}$; (b) $D=12 \
\mathrm{nm}$; (c) $D=14 \ \mathrm{nm}$; (d) $D=16 \ \mathrm{nm}$; (e) $D=18
\ \mathrm{nm}$. }
\label{f4}
\end{figure}

\begin{figure}[tbp]
\includegraphics[width = 4.0in]{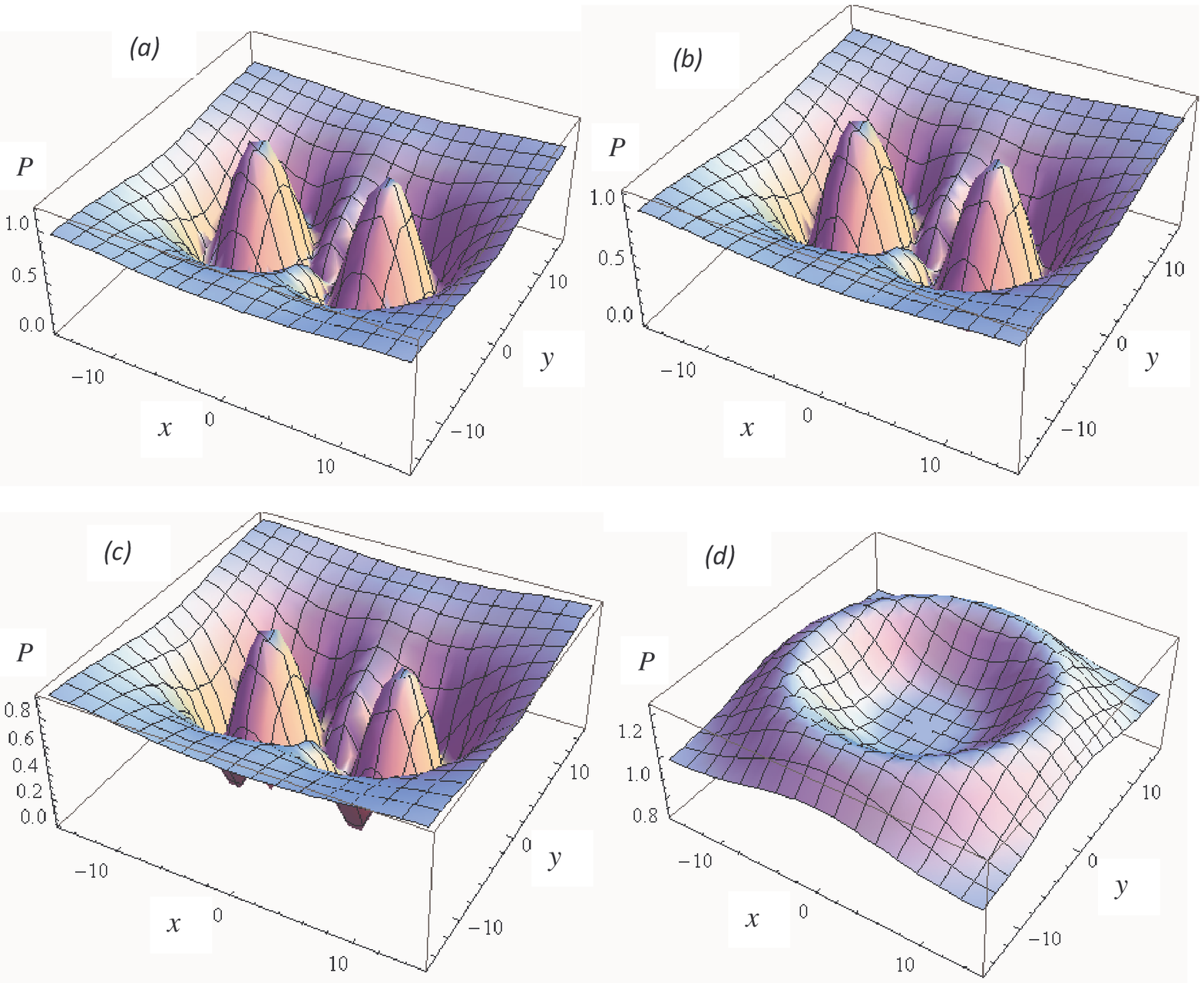}
\caption{ The CPDs for $X^{+}$ trion for the state $n=1$, $l=0$, $N=3$, $L=0$
and different inrewell separations: (a) $D=12 \ \mathrm{nm}$; (b) $D=14 \
\mathrm{nm}$; (c) $D=16 \ \mathrm{nm}$; (d) $D=18 \ \mathrm{nm}$.}
\label{f5}
\end{figure}

\begin{figure}[tbp]
\includegraphics[width = 4.0in]{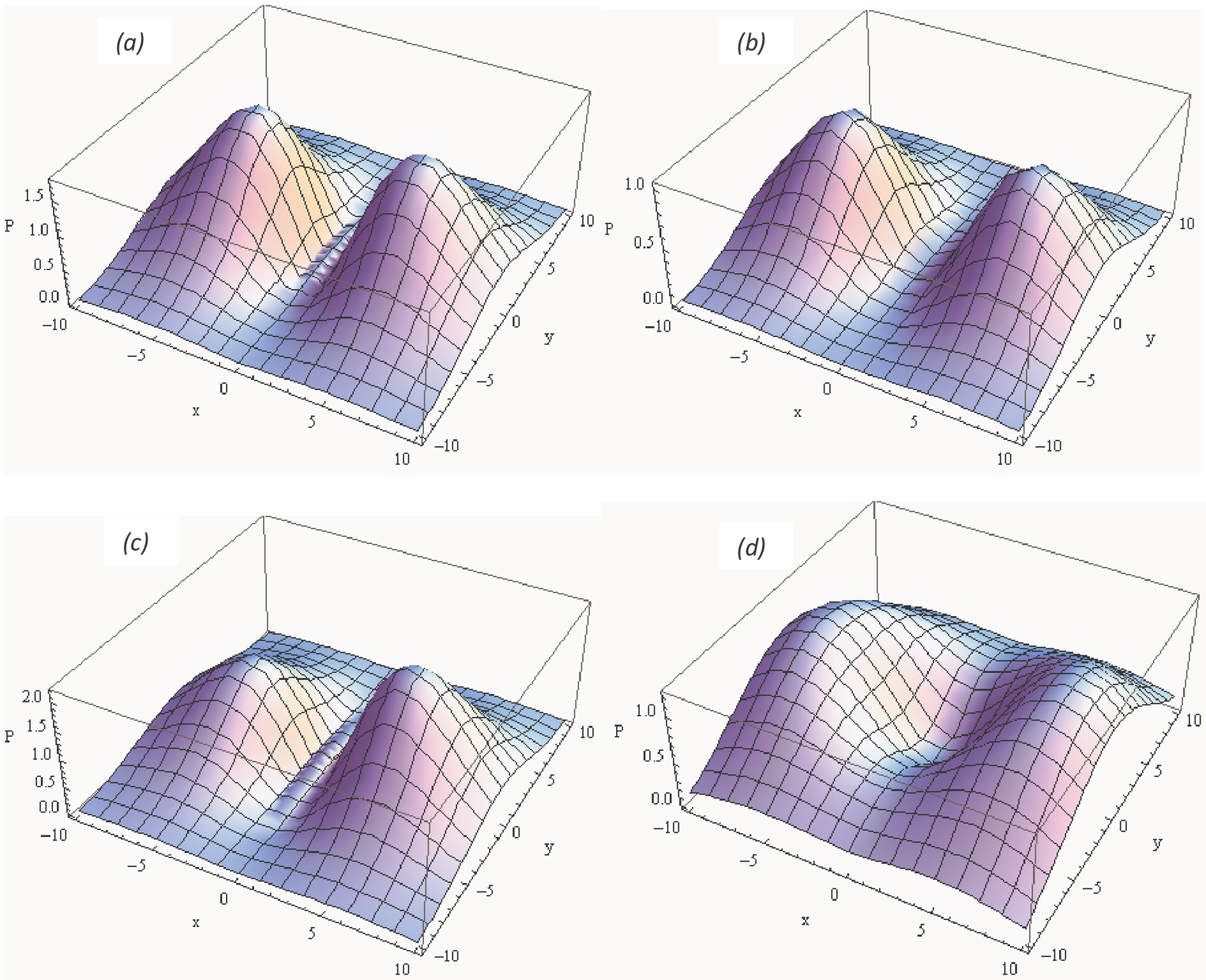}
\caption{The CPDs for $X^{+}$ trion when interwell separation $D=14 \
\mathrm{nm}$ for the states: (a) $n=1$, $l=0$, $N=1$, $L=0$; (b) $n=1$, $l=0$%
, $N=1$, $L=1$; (c) $n=1$, $l=0$, $N=1$, $L=2$; (d) $n=1$, $l=0$, $N=1$, $L=3
$. }
\label{f6}
\end{figure}

The structure of the wave function of the trion which is a product
of the wavefunctions of the exciton and relative motion of exciton
and electron (hole)   can be tested by calculation of the
conditional probability distribution (CPD) which is the pair
correlation function $P(\mathbf{r},\mathbf{r_{0}})$ that expresses
the probability of finding an electron or hole at the position $r$
given when the exciton is located at the position $r_{0}$.
Therefore, the CPD gives us information on when exciton located at
$r_{0}$ sees an electron (hole) at $r$ and defined
as~\cite{Landman1,Landman_prb,Landman2}
\begin{eqnarray}
P(\mathbf{r},\mathbf{r_{0}}) = \frac{\left\langle\Psi\left| \sum_{i\neq
j}^{N}\delta (\mathbf{r}_{i} - \mathbf{r})\delta(\mathbf{r}_{j} - \mathbf{r}%
_{0})\right|\Psi \right\rangle}{\left\langle
\Psi\left|\right.\Psi\right\rangle} \ .  \label{CPD}
\end{eqnarray}
In Eq.~(\ref{CPD}) $\Psi$ is the wave function of the trion, which
is a product of the wavefunctions of the exciton and relative motion
of exciton and electron (hole). We use the accurate analytical wave
functions for the trion to compute the CPDs.  In Fig.~\ref{f4} the
CPDs are displayed for the state $n=1$, $l=0$, $N=2$, $L=0$ for the
different interwell separations that vary from $D=10 \ \mathrm{nm}$
to $D=18 \ \mathrm{nm}$. At the interwell separation $D=10 \
\mathrm{nm}$ the confinement potential is strong and the
contribution of the single-particle energies to the total energy of
the trion is larger. The hole is confined in a rather compact region
so that the CPD does not show clear two peaks. On the contrary, for
$D=14 \ \mathrm{nm}$ the effect of the confinement becomes weaker.
The size of the system grows and we see clearly well separated hump
structure in Fig.~\ref{f4}c and Fig.~\ref{f4}d. For the comparison
in Fig.~\ref{f5} the CPDs for $X^{+}$ trion is presented when the
hole is in the excited state with $N=3$ for different interwell
separations.
The CPDs for four specific states i.e., the $n=1$, $l=0$, $N=1$, $L=0$, the $%
n=1$, $l=0$, $N=1$, $L=1$, the $n=1$, $l=0$, $N=1$, $L=2$ and $n=1$, $l=0$, $%
N=1$, $L=3$ when the interwell separation $D= 14 \ \mathrm{nm}$ in
conjunction with identification of regularities related to the quantum
number $L$ are shown in Fig.~\ref{f6}. By examining the CPD's associated
with selected states one can make a conclusion on the dependence of the CPD
on the quantum number $L$.

\section{Trion Wigner crystal}

\label{Wig}

It is well known that the dilute system of the electrons can form the 2D
Wigner crystal~\cite{Tanatar}. Based on the existence of the trions in the
CQWs we can propose that trions form of 2D Wigner crystal at the low
densities alike to the 2D Wigner crystal formed by the dilute system of the
electrons. The dilute system of trions, when the average distance between
the trions is much larger than the radius of each trion, can be treated as
the dilute system of the fermions with the pair Coulomb repulsion, that
undergo a phase transition and crystallize by forming the Wigner crystal. To
minimize the potential energy, the trions experiencing Coulomb repulsion
form a triangular lattice in 2D system. In other words, this is an example
when Wigner crystal phase is occurring in non-electronic system at low
density.

The Wigner crystal formed by the composite particles was studied in several
recent publications. Ref.~[\onlinecite{Mora}] was devoted to Wigner
crytallization of the dipole cold atoms, while Wigner crystallization of
dipole indirect excitons in QWs was studied in Refs.~[%
\onlinecite{Lozovik_Volkov,Lozovik_prl,Sperlich}]. Wigner crystallization of
the dipole cold atoms in magnetic field was considered in Ref.~[%
\onlinecite{Lozovik_prl}].  The Hamiltonian of the trions in CQWs will be
written alike to the Hamiltonian of composite dipole particles in the papers
listed above, when the energy of the composite particles is counted relative
to the binding energy of the composite dipole particle. The total
Hamiltonian $\hat{H}_{tot}$ per unit area of the interacting system of $%
\tilde{N}$ trions per unit area is similar to that for the electrons~\cite%
{Tanatar}:
\begin{eqnarray}  \label{tot}
\hat{H}_{tot} = \hat{K} + \hat{U} \ .
\end{eqnarray}
In Eq.~(\ref{tot}) $\hat{K}$ is the kinetic energy of the centers mass of
the trions given by
\begin{eqnarray}
\hat{K} = \frac{\hbar^{2}}{2m_{+(-)}S} \sum_{i =1}^{\tilde{N}} \Delta_{%
\mathbf{R}_{i}} \ ,  \label{kin}
\end{eqnarray}%
where $S$ is the area of the system, $M_{-}=m_{e}+m_{h}+m_{e}$ is the mass
of the trion $X^{-}$ and $M_{+}=m_{e}+m_{h}+m_{h}$ is the mass of the trion $%
X^{+}$, and $\hat{U}$ is the potential energy of the trion-trion Coulomb
repulsion per unit area:
\begin{eqnarray}
\hat{U} = \frac{e^{2}}{\epsilon S} \sum_{1\leq i < j\leq \tilde{N}} \frac{1}{%
|\mathbf{R}_{i} - \mathbf{R}_{j}|} \ .  \label{poten}
\end{eqnarray}%
%
In Eq.~(\ref{poten}) $\mathbf{R}_{i}$ and $\mathbf{R}_{j}$ are coordinates
of the centers of the mass of the trions. Let us mention that in the dilute
system of trions when the average distance between the trions is much larger
than the radius of each trion, we can neglect the interaction between
excitons of different trions and exciton in one trion and electron (or hole)
in the other neighbored trion, which are proportional $1/r^{6}$ (see Ref.~[%
\onlinecite{Berman_JETP}]) and $1/r^{4}$ (see Eq.~(\ref{rk3})),
respectively, compare to the Coulomb repulsion between the electrons (or
holes) in the different trions. In this approximation we replace the Coulomb
repulsion between the electrons (or holes) in different trions by the
Coulomb repulsion between the centers of mass of different trions.
Therefore, we can treat the the dilute system of the charged spin-$1/2$
fermions interacting with the pair Coulomb repulsion formed by the trions
the same way as the dilute system of electrons.

According to Ref.~[\onlinecite{Platzman}], the qualitative criterion of the
stability of the Wigner crystal is the condition, when the potential energy
dominates the kinetic energy: $\langle \hat{U}\rangle >\langle \hat{K}
\rangle $, where $\langle \hat{U}\rangle $ is the average potential energy
per unit area, and $\langle \hat{K}\rangle $ is the average value of the
kinetic energies of the centers of mass of trions per unit area. This allows
to determine the density at which the trion system becomes a Wigner crystal.
Let us mention that the averaging of $\hat{U}$ and $\hat{K}$ is calculated
as the averaging by the many-particle antisymmetric wave function of the
centers of mass of trions which is similar to the antisymmetric
many-electron wave function~\cite{Tanatar}.

Let's show that at zero temperature trion Wigner crystal exist at higher
densities than electron (or hole) Wigner crystal. Estimating $\hat{U}$ and $%
\hat{K}$ for the trion system analogously to the electron system~\cite%
{Tanatar}, we conclude that the 2D trion Wigner crystal is stable when the
dimensionless density parameter $r_{s} \geq 37$ at $T=0$. The so-called
Wigner-Seitz radius is defined as  $r_{s} = a/a_{0}$, where $a = (\pi
\rho)^{-1/2}$ is the average distance between the centers of mass of trions,
$\rho$ is the 2D density of trions, and $a_{0} =
\hbar^{2}\epsilon/(M_{+(-)}e^{2})$. The last condition enables us to
determine the density at which the trion gas becomes a Wigner crystal. For
the negative trions $X^{-}$ this condition corresponds to $\rho \leq 4.14
\times 10^{13} \ \mathrm{m}^{-2}$ and for the positive trions $X^{+}$ $\rho
\leq 6.75 \times 10^{13} \ \mathrm{m}^{-2}$. For the electron Wigner crystal
we substitute $a_{0} = \hbar^{2}\epsilon/(m_{e}e^{2})$, which results in $%
\rho \leq 2.41 \times 10^{12} \ \mathrm{m}^{-2}$. For the hole Wigner
crystal we substitute $a_{0} = \hbar^{2}\epsilon/(m_{h}e^{2})$, which
results in $\rho \leq 1.11 \times 10^{13} \ \mathrm{m}^{-2}$. Therefore, the
trion Wigner crystal can be formed at the sufficiently higher densities than
electron (or hole) Wigner crystal, because the mass of trion is greater than
the mass of the electron or hole.

Above we considered the trion gas at zero temperature. At finite
temperatures the defects will always destroy the long-range translational
order in the 2D Wigner crystals. We assume that a 2D Wigner crystal in the
ground state will have no defects. Also let us mentioned that in practice,
it is difficult to experimentally realize a Wigner crystal because quantum
mechanical fluctuations overpower the Coulomb repulsion and quickly cause
disorder. A low dense system is needed.

\section{Conclusions}

\label{conc}

In this Paper we have reduced the three-body restricted problem for a
spatially separated exciton and electron (hole) located in the CQWs to the
2D three-body problem of the exciton and the projection of the electron
(hole) on the plane of the exciton QW. In the limit of large spatial
separation between the electron (hole) and exciton QWs we have obtained the
analytical expression for the wave functions of the trions $X^{-}$ and $X^{+}
$ with the spatially separated electron (hole) and exciton. Analytical
results for the energy spectrum of the trion formed by the exciton and
electron (hole) separated in different quantum wells are obtained, its
dependence on the interwell separations is analyzed and a conditional
probability distribution is calculated. The differences in the binding
energies for $X^{+}$ and $X^{-}$ due to the difference between electron and
hole masses is analyzed. The 2D Wigner crystallization of the trions in the
CQWs is predicted. Due to the fact that mass of the trion is greater than
the mass of electron, the critical density of the formation of the trion
Wigner crystal at zero temperature is sufficiently greater than the critical
density of the electron Wigner crystal.

\acknowledgments

O.~L.~B. and R.~Ya.~K. were supported by PSC CUNY grant 63443-00 41.
O.~L.~B. and R.~Ya.~K. are thankful to Boris Altshuler and Yurii Rubo for
the useful discussions.


\end{document}